\documentclass[preprint,12pt]{elsarticle}

\usepackage{lineno,hyperref}
\usepackage{natbib}
\usepackage{geometry}
\usepackage{fleqn}
\usepackage{graphicx}
\usepackage{newtxtext,newtxmath}
\usepackage{hyperref}
\usepackage{booktabs}
\usepackage{bm}
\usepackage{amsmath}
\usepackage{enumitem}
\usepackage{esvect}
\usepackage[justification=centering]{caption}
\modulolinenumbers[5]

\graphicspath{{./figures/}}
\journal{Elsevier}

\begin{document}

\begin{frontmatter}

\title{\large{Bayesian Inverse Uncertainty Quantification of a MOOSE-based Melt Pool Model for Additive Manufacturing Using Experimental Data}}

\author[NCSU]{Ziyu Xie}

\author[INL1]{Wen Jiang}

\author[INL2]{Congjian Wang}

\author[NCSU]{Xu Wu\corref{mycorrespondingauthor}}
\cortext[mycorrespondingauthor]{Corresponding author}
\ead{xwu27@ncsu.edu}

\address[NCSU]{Department of Nuclear Engineering, North Carolina State University    \\ 
	2500 Stinson Drive, Raleigh, NC 27695 \\}
\address[INL1]{Computational Mechanics \& Materials Department \\
	Idaho National Laboratory, P.O. Box 1625, Idaho Falls, ID 83415 \\}
\address[INL2]{Digital Reactor Technology \& Development Department \\
	Idaho National Laboratory, P.O. Box 1625, Idaho Falls, ID 83415 \\}

\begin{abstract}
Additive manufacturing (AM) technology is being increasingly adopted in a wide variety of application areas due to its ability to rapidly produce, prototype, and customize designs. AM techniques afford significant opportunities in regard to nuclear materials, including an accelerated fabrication process and reduced cost. High-fidelity modeling and simulation (M\&S) of AM processes is being developed in Idaho National Laboratory (INL)'s Multiphysics Object-Oriented Simulation Environment (MOOSE) to support AM process optimization and provide a fundamental understanding of the various physical interactions involved. In this paper, we employ Bayesian inverse uncertainty quantification (UQ) to quantify the input uncertainties in a MOOSE-based melt pool model for AM. Inverse UQ is the process of inversely quantifying the input uncertainties while keeping model predictions consistent with the measurement data. The inverse UQ process takes into account uncertainties from the model, code, and data while simultaneously characterizing the uncertain distributions in the input parameters--rather than merely providing best-fit point estimates. We employ measurement data on melt pool geometry (lengths and depths) to quantify the uncertainties in several melt pool model parameters. Simulation results using the posterior uncertainties have shown improved agreement with experimental data, as compared to those using the prior nominal values. The resulting parameter uncertainties can be used to replace expert opinions in future uncertainty, sensitivity, and validation studies.
\end{abstract}


\end{frontmatter}


\section{Introduction}

Additive manufacturing (AM) technology is being increasingly adopted in a wide variety of application areas due to its ability to rapidly produce, prototype, and customize designs \cite{debroy2018additive}. AM is a broad concept encompassing all technologies that produce parts via a layer-by-layer process. Currently, there are more than 30 AM methods available, and these are primarily differentiated according to fabrication process (e.g., melting or sintering), energy source (e.g., laser or electron beam), material type (e.g., metals or plastics), etc. Below is a list of several representative AM techniques.
\begin{enumerate}[label=(\roman*)]
	\setlength{\itemsep}{0.1pt}
    \item \textit{Fused deposition modeling (FDM)} \cite{thrimurthulu2004optimum}, or \textit{fused filament fabrication} \cite{brenken2018fused}, is a material extrusion technique to continuously deposit filaments of a thermoplastic material in a specific pattern on top of a build platform. Typically, FDM is the most inexpensive AM technique by a wide margin, thus explaining its popularity.
    
    \item \textit{Stereolithography} \cite{bartolo2011stereolithography} uses a laser beam focused on the free surface of a photosensitive liquid in order to induce polymerization of the liquid in that region and transform it into a polymerized solid.
    
    \item \textit{Direct energy deposition (DED)} \cite{gibson2015directed} is also known as \textit{direct laser metal deposition (DLMD)}. In DED processes, a laser is used to fuse materials by melting them together as they are being deposited. Although DED can work for polymers, ceramics, and metal matrix composites, it is predominantly used for metal powders. DED is commonly used for repairing and rebuilding damaged components.
    
    \item \textit{Power bed fusion (PBF)} \cite{tapia2014review} encompasses a group of methods that use a laser or electron beam to melt and fuse powder materials together. Notable subcategories of PBF include (a) \textit{selective heat sintering} \cite{baumers2015selective}, which uses a heated thermal print head to fuse together powder materials, with new layers added via a roller in between each layer-fusion process; (b) \textit{selective laser sintering (SLS)} \cite{kruth2005binding}, which uses a high-power laser to fuse together small-particle plastic, metal, ceramic, or glass powders into a mass featuring the desired 3-D shape; (c) \textit{selective laser melting (SLM)} \cite{yap2015review}, also called direct metal laser melting and laser powder bed fusion \cite{king2015laser}, which uses laser to melt and fuse together metallic powders, and (d) \textit{Electron beam melting (EBM)} \cite{zah2010modelling}, also called selective electron beam melting, is similar to SLM but uses an electron beam as the energy source.
\end{enumerate}

To facilitate the application of advanced AM techniques for producing materials used in high-consequence systems such as nuclear reactors, high-fidelity modeling and simulation (M\&S) are needed, in addition to extensive experimental tests. Numerous computational models have been developed that involve multiscale, multiphysics AM phenomena. King et al. \cite{king2015laser} reviewed the physics, computational, and materials challenges stemming from the broad range of length/time scales and temperature ranges associated with laser beam PBF. Francois et al. \cite{francois2017modeling} reviewed the challenges and opportunities related to M\&S of AM processes for metals, and the predictive representation of their mechanical performance at different scales. A new 3-D mathematical model of laser powder deposition was developed in \cite{fathi2006prediction} to predict the temperature field, melt pool depth, and dilution. The evolution of temperature and velocity fields during laser spot welding of 304 stainless steel was studied in \cite{he2003heat} \cite{he2003alloying} using a transient heat transfer and fluid flow model based on the solution of the equations of conservation of mass, momentum, and energy in the weld pool. Morville et al. \cite{morville20122d} developed a 2-D transient DLMD model that included filler material with surface tension, as well as thermocapillary effects at the free surface. The dynamic shape of the molten zone is explicitly described by a moving mesh based on an arbitrary Lagrangian-Eulerian method. This model was used to analyze the influence of the process parameters (e.g., laser power, scanning speed, and powder feed rate) on the melt pool behavior.

As more experimental and computational data become available for various AM systems, data-science-based methodologies such as data mining and machine learning (ML) are being used to improve our understanding of the physical phenomena involved in AM processes. Kamath \cite{kamath2016data} demonstrated that techniques from data mining and statistics (e.g., feature selection, improved sampling, and data-driven surrogate modeling) can complement those from physical modeling in order to provide greater insight into complex processes such as SLM. Khanzadeh et al. \cite{khanzadeh2018porosity} investigated the relationship between melt pool characteristics and defect (porosity) occurrence in an as-built AM part. Supervised ML methods were utilized to identify the patterns of melt pool images and build a black box model for the probability distribution of class labels (namely, porosity), based on data of the melt pool characteristics. Francis and Bian \cite{francis2019deep} developed a novel deep learning (DL) approach for accurately predicting geometric distortions in AM products by analyzing big data via a host of sensors. Kapusuzoglu and Mahadevan \cite{kapusuzoglu2020physics} investigated several physics-informed and hybrid ML strategies that incorporate physics knowledge in experimental, data-driven, DL models for predicting the bond quality and porosity of FDM parts.

Even though M\&S of various AM processes has tremendously advanced over the last two decades, it remains crucial to determine the accuracy of these predictions by using concrete, quantifiable measurements of the M\&S uncertainties. Predictive simulation must involve a systematic treatment of model/data uncertainties and their propagation through a computational model in order to produce predictions of quantities-of-interest (QoIs) with quantified uncertainty. Much work has already been done on forward uncertainty quantification (UQ) of AM models, the process for obtaining the uncertainties in QoIs by propagating the uncertainties in input parameters through the computer model. Lopez et al. \cite{lopez2016identifying} discussed the origin and propagation of uncertainties in laser beam PBF models. Four sources of uncertainty were identified: modeling assumptions, unknown simulation parameters, numerical approximations, and measurement errors in calibration data. Corresponding UQ methods were illustrated through a case study in which the prediction uncertainty for the melt pool width was quantified using a thermal model. Hu and Mahadevan \cite{hu2017uncertainty2} reviewed the current status, needs, and opportunities pertaining to UQ and management of AM processes. In another work \cite{hu2017uncertainty1}, the same authors' discussion on UQ for AM focused on the prediction of material properties. Nath et al. \cite{nath2019uncertainty} presented a systematic UQ framework for quantifying the grain morphology uncertainty caused by various sources of uncertainty in the DLMD simulation process.

Surrogate modeling has also been widely used to reduce the computational cost in forward UQ of AM models. For example, Vohra et al. \cite{vohra2020fast} presented a novel approach aimed at combining dimension reduction in the output space (with principal component analysis) with dimension reduction in the input space (with active subspace). A surrogate model was constructed to map a sparse set of representative random variables (from random field discretization) to a set of input variables in a low-dimensional subspace. The surrogate model was then used to quantify the uncertainty in residual stress in an EBM-fabricated component caused by the stochastic nature of the process variables and material properties. Wang et al. \cite{wang2019data} \cite{wang2019uncertainty} developed a physics-informed, data-driven modeling framework in which multilevel data-driven surrogate models are constructed based on extensive multiscale, multiphysics AM models. The proposed data-driven UQ framework was used to predict a process-structure-property relation for EBM of Ti-6Al-4V material. Another example of surrogate modeling is found in \cite{tapia2018uncertainty}, with the authors employing the generalized polynomial chaos expansions for forward UQ of the melt pool width predicted by two laser beam PBF thermal models.

A typical challenge in forward UQ is that it requires input uncertainty information, usually generated through ``expert opinion'' or ``user self-assessment.'' Such ad hoc characterization lacks mathematical rigor and can be subjective, leading to biased forward UQ results. Inverse UQ is the process of inversely quantifying the input uncertainties based on model fits to measured data. Even though the need to obtain input parameter uncertainties via inverse UQ (or similarly, statistical calibration) was acknowledged in previous works \cite{wang2019uncertainty} \cite{olleak2020calibration}, work on inverse UQ of AM models remains very limited. Recently, an innovative inverse UQ method based on the modular Bayesian approach \cite{wu2018inverse-part1} was developed to calibrate the parameters in computational models. It was successfully applied to the system thermal-hydraulics code TRACE \cite{wu2018inverse-part2} and the fuel performance code BISON \cite{wu2018kriging}. In this paper, this Bayesian inverse UQ approach is applied to a melt pool model based on experimental data for melt pool length and depth.

High-fidelity M\&S of AM processes is being performed to support optimal AM process development and provide a fundamental understanding of the various physical interactions involved. A multiphysics melt pool model  has been developed using Multiphysics Object-Oriented Simulation Environment (MOOSE) framework \cite{permann2020moose} to analysis highly nonlinear melt pool dynamics. However, such a model is subject to uncertainties caused by (1) uncertain model parameters, (2) inaccurate/insufficient underlying physics compared to the real phenomena, and (3) numerical approximation errors, etc.

In this paper, we will perform Bayesian inverse UQ to quantify the parameter uncertainties in the MOOSE-based melt pool model developed at INL. Under the Bayesian framework, inverse UQ tries to build a relation among uncertain inputs, model prediction, and experimental data by using the ``model updating equation'' to formulate posterior distributions for the uncertainty inputs while simultaneously considering uncertainties from other sources. The quantified parameter uncertainties can be used for future uncertainty, sensitivity, and validation studies. Section 2 briefly discuss the MOOSE-based melt pool model. Section 3 presents the general methodology of the Bayesian inverse UQ method. The results for inverse UQ are included in Section 4. The paper is concluded in Section 5.

\section{The MOOSE-based Melt Pool Model}

During a laser-based AM process, a high-intensity moving energy source strikes the metal powders. The powders absorb energy to change their material phases and will form a melt pool locally. ``Melt pool" refers to the region at the laser-powder interface where powder particles melt to form a pool of molten metal. Correctly predicting the major melt pool characteristics (e.g., geometry and temperature) is crucial, since these indicate how well adjacent tracks and successive layers will bond to one another. The heat will transfer via convection/conduction and create non-uniform temperature profiles. The melt pool physics is categorized as a nonlinear, non-equilibrium multiphysics process. For computational efficiency, powder particles are simplified as a homogenized continuum medium, and a continuum finite element model (FEM) was developed in MOOSE \cite{permann2020moose} \cite{peterson2018overview} to describe the relevant multiphysics phenomena, including generation of the powder layer, melting and solidification, melt pool dynamics, and the thermal-capillary, buoyant, conductive, and convective heat transport processes.

The level set method is a numerical approach for tracking the free interface in melt pool modeling. In this method, the location of the moving interface is tied to an isocontour of a scale field. The mesh is fixed in time, and the material moves through the mesh, making this technique suitable for severe interface deformations and topology changes. In this work, we use a conservative level set method \cite{he2003alloying} \cite{wen2010modeling} \cite{courtois2014complete} to accurately model the evolution of the liquid-gas interface. Table \ref{table:table1-MP-Model-Symbols} lists the symbols used in the model, along with their physical meanings and units. The level set evolution is written as:
\begin{equation}    \label{equation:MP1-Level-Set}
	\frac{\partial \phi(\vv{x},t)}{\partial t} + \vv{u}(\vv{x},t) \cdot \nabla \phi (\vv{x},t) + F_p | \nabla \phi(\vv{x},t) | = 0
\end{equation}
where $\phi (\vv{x},t)$ is the level set function $\phi (\vv{x},t) = \pm d$, where $d$ is the actual distance from the zero level set and the $\pm$ sign represents the outside/inside region of the interface. $\vv{u}(\vv{x},t)$ is the fluid velocity. $F_p$ is the powder addition speed.

\begin{table}[ht]
	\centering
	\footnotesize
	\caption{Physical meanings and units of the symbols used in the MOOSE-based melt pool model.}
	\label{table:table1-MP-Model-Symbols} 
	\begin{tabular}{l c c }
		\toprule
		Parameters  &  Unit  &  Symbol   \\ 
		\midrule
		Level set function              &  $\mathrm{ m }$                          & $\phi(\vv{x},t)$  \\
		Fluid velocity                  &  $\mathrm{m / s}$                        & $\vv{u}(\vv{x},t)$ \\
		Powder addition speed           &  $\mathrm{m / s}$                        & $F_p$             \\
		Density                         &  $\mathrm{kg / m^3}$                     & $\rho$            \\
		Enthalpy                        &  $\mathrm{J / ( kg )}$                   & $h$               \\
		Thermal conductivity            &  $\mathrm{J / ( m \cdot s \cdot K )}$    & $k$               \\
		Dynamic viscosity               &  $\mathrm{kg / ( m \cdot s )}$           & $\mu$             \\
		Laser power                     &  $\mathrm{J/s}$                          & $P$               \\
		Effective beam radius           &  $\mathrm{m}$                            & $R_b$             \\
		Laser energy absorption coefficient     &  $-$                             & $\alpha$          \\
		Heat transfer coefficient       &  $\mathrm{J / (s \cdot m^2 \cdot K)}$    & $A_h$             \\
		Stefan-Boltzmann constant       &  $\mathrm{J / (s \cdot m^2 \cdot K^4)}$  & $\sigma$          \\
		Material emissivity             &  $-$                                     & $\epsilon$        \\
        Melt pool temperature           &  $\mathrm{K}$                            & $T$               \\
        Ambient temperature             &  $\mathrm{K}$                            & $T_0$             \\
        Thermal expansion coefficient   &  $-$                                     & $\beta_l$         \\
        Gravity vector                  &  $\mathrm{m / s^2} $                     & $\vv{g}$          \\
        Reference temperature           &  $\mathrm{K} $                           & $T_r$             \\    
        Isotropic permeability          &  $\mathrm{m^2} $                         & $K$               \\
        Surface tension coefficient     &  $\mathrm{N / m} $                       & $\gamma$          \\
        Surface curvature               &  $\mathrm{m^{-1}} $                      & $\kappa$          \\
        Normal vector to the free surface      &  $-$                              & $\vv{n}$          \\
        Thermal capillary coefficient   &  $\mathrm{N / ( m \cdot K)} $            & $\gamma_T$        \\
		\bottomrule
	\end{tabular}
\end{table}

The material properties of density $\rho$, enthalpy $h$, thermal conductivity $k$, and dynamic viscosity $\mu$ are smoothly varied across the interface between gas and solid/liquid using a level set variable.  The solid-liquid region of the metal is described as pure solid, pure liquid, or a solid-liquid mixture (mushy zone) in which the material properties are determined via the mass and volume fractions.

A continuum FEM is used to describe the relevant multiphysics phenomena, including the generation of the powder layer, melting and solidification, melt pool dynamics, and the thermal-capillary, buoyant, conductive, and convective heat transport processes. The following conservation equations of mass (Equation \ref{equation:MP2-Mass}), energy (Equation \ref{equation:MP3-Energy}), and momentum (Equation \ref{equation:MP4-Momentum}) are solved in a fully coupled manner. Since the gas and liquid flows are assumed incompressible, the mass conservation equation simplifies to:
\begin{equation}    \label{equation:MP2-Mass}
	\nabla \cdot \vv{u} = 0
\end{equation}

The energy conservation equation is described by:
\begin{multline}    \label{equation:MP3-Energy}
	\rho \frac{\partial h}{\partial t} + \rho \nabla \cdot (\vv{u} h) = 
	\nabla \cdot (k \nabla T)  \\
	+  \frac{2P \alpha}{\pi R_b^2} \exp \left( \frac{-2r^2}{R_b^2} \right) |\nabla \phi|  -  A_h (T - T_0) | \nabla \phi | - \sigma \epsilon (T^4 - T_0^4) | \nabla \phi |
\end{multline}
where the last three terms on the right-hand side represent heat flux from the laser, heat loss through convection, and heat loss through radiation, respectively. In these terms, $P$ is the laser power, $R_b$ is the effective beam radius, $\alpha$ is the laser energy absorption coefficient, $A_h$ is the heat transfer coefficient, $\sigma$ is the Stefan-Boltzmann constant, $\epsilon$ is the material emissivity, and $T_0$ is the ambient temperature. The momentum equation is expressed by:
\begin{multline}    \label{equation:MP4-Momentum}
	\rho \left( \frac{\partial{\vv u} }{\partial t}  +  \vv{u} \cdot {\nabla\vv u} \right)  =  
	\nabla \left[ -p \mathbf{I} + \mu \left( \nabla \vv{u} + \nabla \vv{u}^T \right) \right]    \\
	-  \rho_l \beta_l \left( T - T_r \right) \vv{g}  -  \frac{\mu_m}{K} \vv{u} + \gamma \vv{n} \kappa | \nabla \phi | - \gamma_T \nabla_s T | \nabla \phi |
\end{multline}
where the last four terms on the right-hand side represent the buoyancy, Darcy damping, capillary, and thermal-capillary (Marangoni) forces respectively. In these terms, $\beta_l$ is the thermal expansion coefficient, $\vv g$ is the gravity vector, $T_r$ is the reference temperature, $K$ is the isotropic permeability, $\gamma$ is the surface tension coefficient, $\kappa$ is the surface curvature, $\vv n$ is the normal vector to the free surface, $\gamma_T$ is the thermal capillary coefficient, and $\nabla_s$ is the surface gradient operator.

\begin{figure}[htbp]
	\centering
	\includegraphics[width=0.6\textwidth]{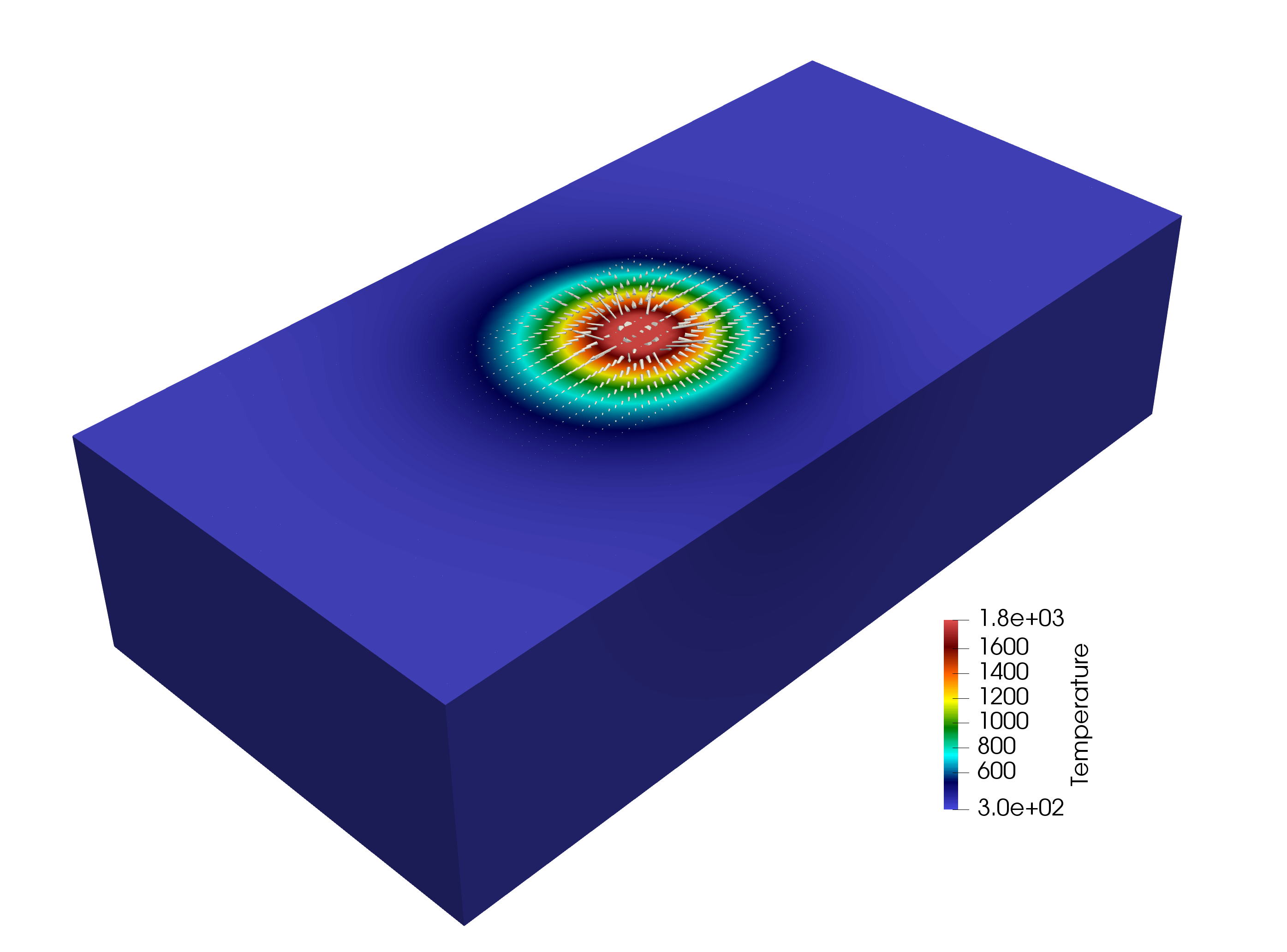}
	\caption[]{Melt pool predictions using MOOSE.}
	\label{figure:fig1-Melt-Pool-Illustration}
\end{figure}

Equations (\ref{equation:MP1-Level-Set})--(\ref{equation:MP4-Momentum}) are implemented in MOOSE using the automatic differentiation capability \cite{lindsay2021automatic}  and solved in a fully implicit manner. Figure \ref{figure:fig1-Melt-Pool-Illustration} shows 3D melt pool predictions using the MOOSE. For Bayesian inverse UQ, we made an simplification by assuming that the level set is not evolving and that the gas-liquid interface remains flat. Due to the experimental data available, we will only focus on the melt pool geometry (length and depth). Inverse UQ using melt pool temperature distributions will be performed in the future, once the measurement data are available.

\section{Bayesian Inverse UQ Methodology}

Forward uncertainty propagation requires knowledge of the computer model input uncertainty information; for example, the probability distribution function (PDF), mean value, standard deviation, lower/upper bounds, etc. Historically, ``expert opinion'' or ``user self-evaluation'' were widely used to generate such information, a subjective process that is not rigorous mathematically. Different users may use different parameter uncertainties for the same problem within the same code, leading to inconsistencies. Bayesian inverse UQ \cite{wu2018inverse-part1} is a process for quantifying the uncertainties in random input parameters that are consistent with experimental data. Within the Bayesian framework, it seeks the posterior distributions rather than point estimates of ``best-fit'' values of the uncertain input parameters.

Consider the computer model $\mathbf{y}^{\text{M}} = \mathbf{y}^{\text{M}} \left( \mathbf{x}, \bm{\theta} \right)$, where $\mathbf{y}^{\text{M}}$ is the model response, $\mathbf{x}$ is the vector of design variables, and $\bm{\theta}$ is the vector of calibration parameters. See \cite{wu2018inverse-part1} for a detailed discussion on the classification of input parameters as design variables and calibration parameters. Inverse UQ is based on the ``model updating formulation'':
\begin{equation}	 \label{equation:IUQ2-Model-Update-Eqn}
	\mathbf{y}^{\text{E}} (\mathbf{x}) = \mathbf{y}^{\text{M}} \left( \mathbf{x}, \bm{\theta}^{*} \right) + \delta(\mathbf{x}) + \bm{\epsilon}
\end{equation}
where $\delta(\mathbf{x})$ is the model discrepancy, also called model uncertainty, model inadequacy, or model bias/error \cite{kennedy2001bayesian}. It results from incomplete or inaccurate underlying physics, numerical approximation errors, and/or other inaccuracies that would exist even if all the parameters in the computer model were known. $\mathbf{y}^{\text{E}} (\mathbf{x})$ represents experimental observations, while $\bm{\epsilon} \sim \mathcal{N} \left( 0, \mathbf{\Sigma}_{\text{exp}} \right)$ represents the measurement error/uncertainty usually treated as a zero-mean Gaussian. $\bm{\theta}^{*}$ is the unknown ``true'' value of $\bm{\theta}$. The goal of inverse UQ is to find the distribution of $\bm{\theta}^{*}$, since the exact value can never be learned, due to the limited data available.

Based on the model updating formulation and the Gaussian assumption of $\bm{\epsilon}$, $\bm{\epsilon} = \mathbf{y}^{\text{E}} (\mathbf{x}) - \mathbf{y}^{\text{M}} \left( \mathbf{x}, \bm{\theta}^{*} \right) - \delta(\mathbf{x})$ follows a multi-dimensional Gaussian distribution. The posterior can be written as:
\begin{multline}	 \label{equation:IUQ3-Posterior}
	p \left( \bm{\theta}^{*} | \mathbf{y}^{\text{E}}, \mathbf{y}^{\text{M}}\right)  \propto  p \left( \bm{\theta}^{*} \right) \cdot p \left( \mathbf{y}^{\text{E}}, \mathbf{y}^{\text{M}} | \bm{\theta}^{*} \right) 	\\
	\propto  p \left( \bm{\theta}^{*} \right) \cdot \frac{1}{\sqrt{|\mathbf{\Sigma}|}}  \cdot \text{exp} \left[  - \frac{1}{2} \left[ \mathbf{y}^{\text{E}} - \mathbf{y}^{\text{M}} - \delta \right]^\top \mathbf{\Sigma}^{-1} \left[ \mathbf{y}^{\text{E}} - \mathbf{y}^{\text{M}} - \delta \right] \right]
\end{multline}
where $p \left( \bm{\theta}^{*} \right)$ is the prior and $p \left( \mathbf{y}^{\text{E}}, \mathbf{y}^{\text{M}} | \bm{\theta}^{*} \right)$ is the likelihood function. Prior and  posterior probabilities represent degrees of belief about possible values of $\bm{\theta}^{*}$, before and after observing the experimental data. $\mathbf{\Sigma}$ is the covariance of the likelihood, consisting of three parts:
\begin{equation}
	\mathbf{\Sigma} = \mathbf{\Sigma}_{\text{exp}} + \mathbf{\Sigma}_{\text{bias}} + \mathbf{\Sigma}_{\text{code}}
\end{equation}

The first part, $\mathbf{\Sigma}_{\text{exp}}$, is the experimental uncertainty due to measurement error. The second part, $\mathbf{\Sigma}_{\text{bias}}$, represents the model uncertainty due to missing/inaccurate underlying physics and numerical approximation errors. The third term, $\mathbf{\Sigma}_{\text{code}}$, is called code uncertainty (or interpolation uncertainty) because we do not know the computer code outputs for every input, especially when the code is computationally prohibitive. In this case, we might choose to use a metamodel, also called a surrogate model. Note that $\mathbf{\Sigma}_{\text{code}} = \mathbf{0}$ if the computer model is used rather than its surrogates.

Equation (\ref{equation:IUQ3-Posterior}) shows that Bayesian inverse UQ uses a very comprehensive formulation that accounts for all major quantifiable uncertainties in M\&S (i.e., parameter, model, experiment, and code uncertainties). Note that numerical uncertainty, or numerical approximation error, is usually treated as a component of the model uncertainty $\mathbf{\Sigma}_{\text{bias}}$. The mathematical description of the model uncertainty $\mathbf{\Sigma}_{\text{bias}}$ is especially challenging in inverse UQ. See \cite{wu2018inverse-part1} for a solution based on the modular Bayesian approach. Most available methods require a considerable amount of data to ``train'' a statistical/mathematical model for $\mathbf{\Sigma}_{\text{bias}}$. However, experimental data for melt pool geometry are very limited. Therefore, in this work, the model uncertainty is not considered.

With the posterior PDF in Equation (\ref{equation:IUQ3-Posterior}), Bayesian inverse UQ then uses Markov Chain Monte Carlo (MCMC) sampling \cite{andrieu2008tutorial} \cite{brooks2011handbook} to generate samples following a probability density proportional to the posterior PDF, without knowing the normalizing constant. Because MCMC typically requires $\sim$10,000 samples, which is expensive for computationally prohibitive models, we plan to use metamodels to speed up the sampling. Metamodels (also called surrogate models, response surfaces, or emulators) are approximations of the input/output relation of a computer model. Metamodels are built from a limited number of original model runs (training sets) and a learning algorithm, and they can be evaluated very quickly. In this paper, we use a Gaussian process (GP) surrogate model \cite{williams1995gaussian} \cite{santner2003design} for the MOOSE-based melt pool model. This is primarily because GP is unique in that it not only provides a mean estimator at any untried input location, but also the corresponding variance estimator (mean squared error), which can serve as the code uncertainty $\mathbf{\Sigma}_{\text{code}}$. In this way, during MCMC sampling, we do not become overconfident in the GP surrogate model, since we have a code uncertainty term in the variance of the posterior PDF.

\section{APPLICATION}

Table \ref{table:table2-list-Calibration-Parameters} shows the MOOSE-based melt pool model's uncertain input parameters, which the model developers suggest can potentially affect melt pool geometry predictions. The prior distributions of these parameters are assumed to be uniform, and their lower/upper bounds (multiplication factors) are shown in Table \ref{table:table2-list-Calibration-Parameters}. The melt pool material is 304 stainless steel. The nominal values were determined based on \cite{he2003alloying}. The prior bounds were selected based on suggestions from the model developers. Most of the ranges are not very wide, due to consideration of their physical meanings.

\begin{table}[ht]
	\centering
	\footnotesize
	\caption{List of uncertain input parameters for inverse UQ.}
	\label{table:table2-list-Calibration-Parameters} 
	\begin{tabular}{l c c c c}
		\toprule
		Parameters  &  Unit  &  Symbol  &  Nominal values  &  Bounds \\ 
		\midrule
		Laser energy absorption coefficient &  $-$                                  & $\alpha$   & 0.27   & [0.5, 1.5]  \\
		Heat transfer coefficient       &  $\mathrm{J / ( m^2 \cdot s \cdot K )}$  & $A_h$      & 100    & [0.8, 1.2]  \\
		Material emissivity             &  $-$                                     & $\epsilon$ & 0.59   & [0.5, 1.5]  \\
		Specific heat                   &  $\mathrm{J / ( kg \cdot K )}$           & $c_l$      & 837.4  & [0.9, 1.1]  \\
		Effective thermal conductivity  &  $\mathrm{J / ( m \cdot s \cdot K )}$    & $k_l$      & 209.3  & [0.9, 1.1]  \\
		Latent heat                     &  $\mathrm{J / kg}$                       & $L$        & $2.5{\times}10^5$  & [0.9, 1.1]  \\
		Effective viscosity             &  $\mathrm{kg / ( m \cdot s )}$           & $\mu_l$    & 0.1    & [0.5, 2.0] \\
		Thermal capillary               &  $\mathrm{N / ( m \cdot K )}$            & $\gamma_T$ & $-4.3{\times}10^{-4}$  & [0.9, 1.1]  \\
		\bottomrule
	\end{tabular}
\end{table}

In the following sections, we first run the model with all input parameters fixed at the prior nominal values, then compare the simulation results with the selected experimental data. We then build a GP surrogate model for the MOOSE-based melt pool model and validate its accuracy in replacing the original model. Once the GP surrogate model is proven accurate enough, we will use it in MCMC sampling to explore the posterior PDF. Lastly, we will run the model at the inversely quantified posterior mean values and compare the melt pool geometry predictions with the experimental data to determine any improvement in agreement. In this manner, inverse UQ will improve the predictive capability of the MOOSE-based melt pool model through a rigorous calibration process.

\subsection{Comparison between experimental data and the simulation at prior nominal values}

\begin{figure}[ht]
	\centering
	\includegraphics[width=0.85\textwidth]{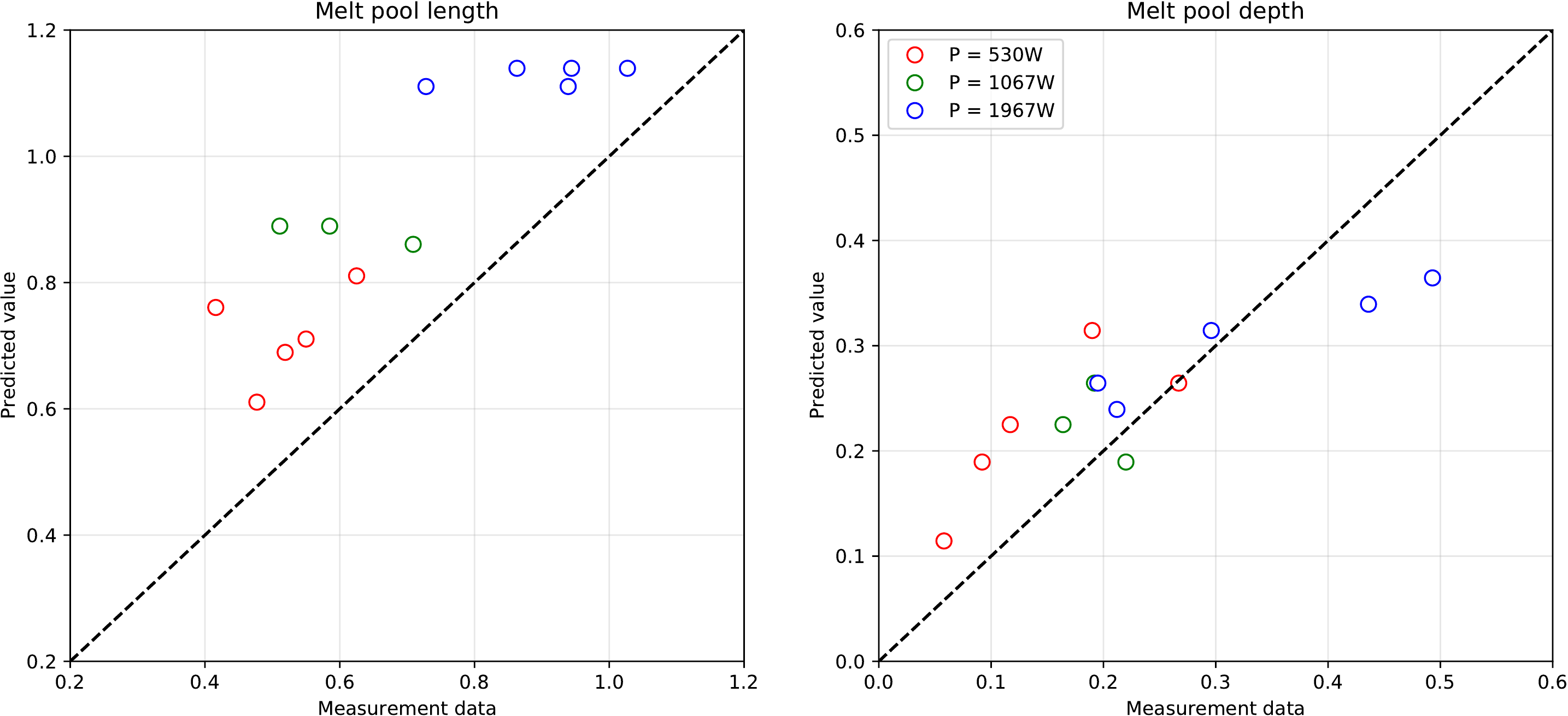}
	\caption[]{Comparison between the simulation and experimental data for melt pool sizes.}
	\label{figure:fig2-Data-vs-Simulation-Prior}
\end{figure}

Inverse UQ requires high-quality experimental data to quantify the uncertainties in the input parameters. Experimental data for melt pool geometry (length and depth) are very limited. In this work, we use the melt pool size measurement data from \cite{he2003heat} and \cite{he2003alloying} as the physical data for inverse UQ. The experiments presented in \cite{he2003heat} and \cite{he2003alloying} were performed under 13 different conditions, as defined by three design variables: power, laser beam radius, and laser pulse duration. The available data and their corresponding design variables are presented in Table \ref{table:table3-Design-Variables-and-Exp-Data}. The MOOSE-based melt pool model runs at prior nominal values of the uncertain inputs in Table \ref{table:table2-list-Calibration-Parameters}, producing simulated melt pool sizes that correspond to the design variables in Table \ref{table:table3-Design-Variables-and-Exp-Data}. A comparison between the simulation and experimental data is shown in Figure \ref{figure:fig2-Data-vs-Simulation-Prior}. It can be seen that the model has relatively good performance for melt pool depth, but consistently overpredicts melt pool length.

\begin{table}[ht]
	\centering
	\footnotesize
	\caption{Experimental data for melt pool sizes, along with the corresponding design variables.}
	\label{table:table3-Design-Variables-and-Exp-Data} 
	\begin{tabular}{cccccc}
		\toprule 
		Index  &  Power  &  Laser beam  &  Laser pulse  &  Melt pool  &  Melt pool   \\
		       &  (W)    & radius (mm)  & duration (ms) & length (mm) &  depth (mm)  \\
		\midrule
		1  & 530  &  0.159  & 4    & 0.625 & 0.190 \\ 
		2  & 530  &  0.210  & 4    & 0.416 & 0.267 \\  
		3  & 530  &  0.272  & 4    & 0.550 & 0.117 \\ 
		4  & 530  &  0.313  & 4    & 0.519 & 0.092 \\  
		5  & 530  &  0.433  & 4    & 0.477 & 0.058 \\ 
		6  & 1067 &  0.325  & 3    & 0.585 & 0.192 \\ 
		7  & 1067 &  0.389  & 3    & 0.511 & 0.164 \\ 
		8  & 1067 &  0.466  & 3    & 0.709 & 0.220 \\ 
		9  & 1967 &  0.350  & 3    & 0.939 & 0.493 \\ 
		10 & 1967 &  0.379  & 3    & 0.728 & 0.436 \\ 
		11 & 1967 &  0.428  & 3    & 0.944 & 0.296 \\ 
		12 & 1967 &  0.521  & 3    & 0.863 & 0.195 \\ 
		13 & 1967 &  0.570  & 3    & 1.027 & 0.212 \\
		\bottomrule
	\end{tabular}
\end{table}

\subsection{Building the GP surrogate model}

The MOOSE-based melt pool model, being an FEM, is very computationally expensive, even with INL's high-performance computing platform. As mentioned, a typical MCMC sampling requires $\sim$10,000 samples for a good exploration of the posterior PDF. To alleviate the computational burden, we chose to build a GP surrogate model. GP has a unique feature lacking in other ML methods such as artificial neural networks; namely, it directly provides the variance (or mean squared error) of the estimation, which then serves as the code uncertainty $\mathbf{\Sigma}_{\text{code}}$.

\begin{figure}[htb]
	\centering
	\includegraphics[width=0.85\textwidth]{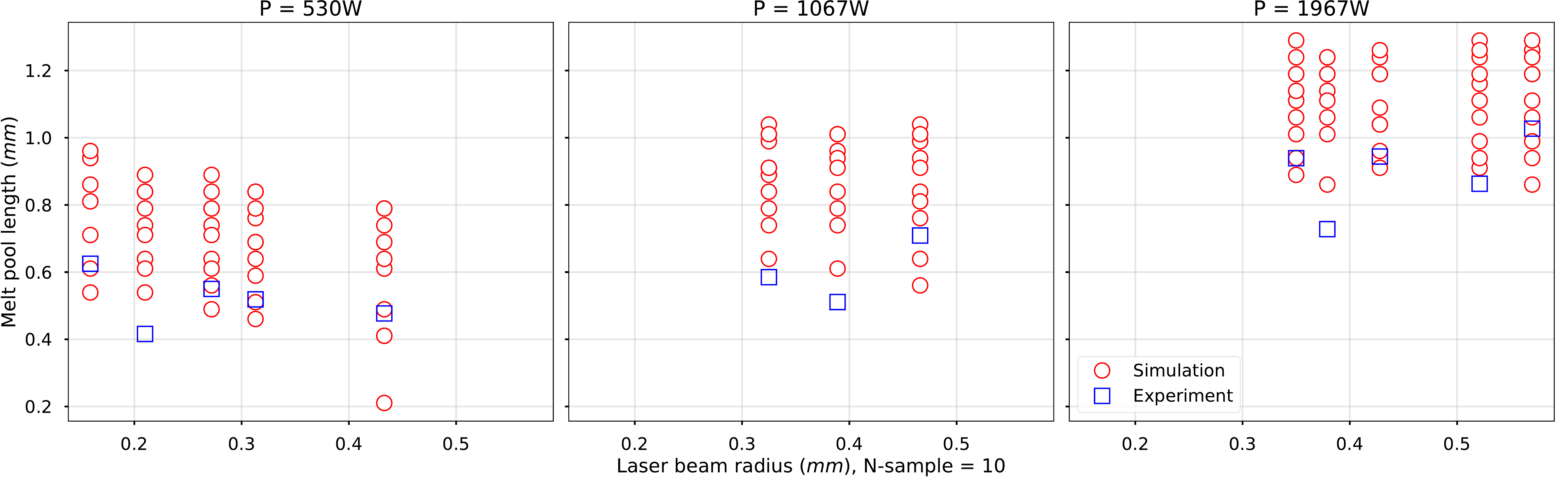}
	\caption[]{Model outputs for melt pool length, with 10 training samples for each experiment.}
	\label{figure:fig4-DoE10-Melt-Pool-Length}
\end{figure}

\begin{figure}[htb]
	\centering
	\includegraphics[width=0.85\textwidth]{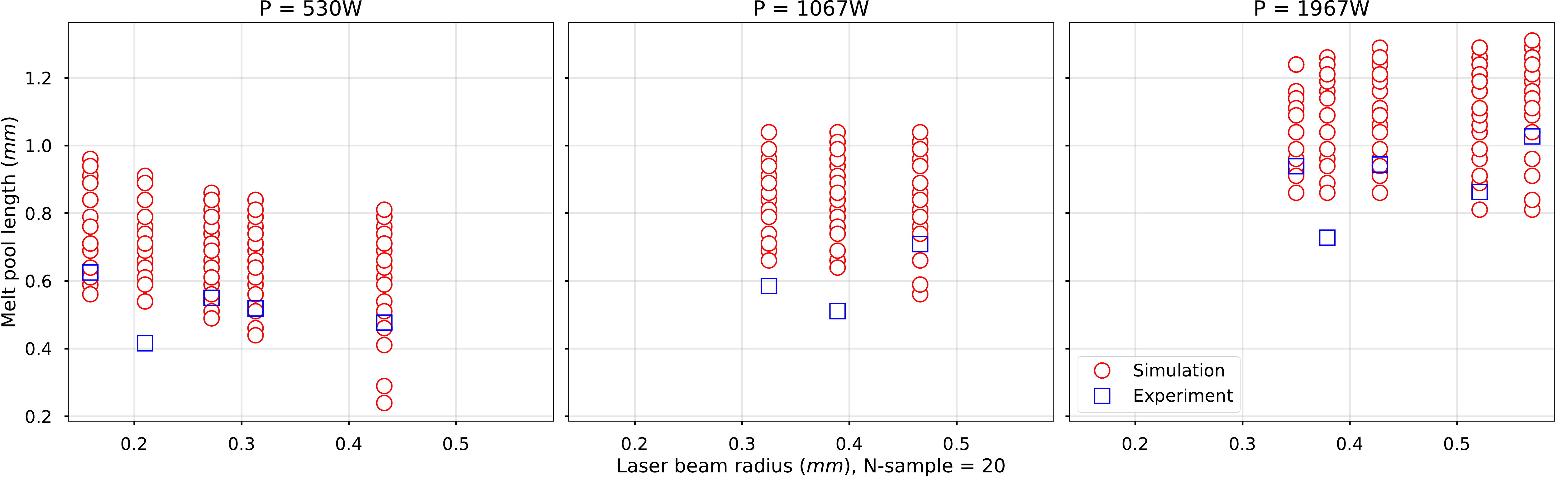}
	\caption[]{Model outputs for melt pool length, with 20 training samples for each experiment.}
	\label{figure:fig4-DoE20-Melt-Pool-Length}
\end{figure}

As the GP surrogate model will replace the original simulation model during MCMC sampling, its accuracy must be validated. Generally, the more training samples generated by the original model, the higher the accuracy of the GP surrogate model. However, this contradicts the intention behind using the surrogate model. To reduce the number of training samples, we used Latin hypercube sampling \cite{helton2003latin} to generate training samples in accordance with the prior space defined by Table \ref{table:table2-list-Calibration-Parameters}, as it has better coverage of the parameter space than crude Monte Carlo sampling. Because the number of samples sufficient to build a surrogate model is unknown, we generated 5, 10, 15, 20, and 25 samples for each of the 13 experimental conditions in Table \ref{table:table3-Design-Variables-and-Exp-Data}, then ran the melt pool model for these samples to generate the melt pool size predictions. As an example, Figures \ref{figure:fig4-DoE10-Melt-Pool-Length} and \ref{figure:fig4-DoE20-Melt-Pool-Length} compare training outputs with the experimental data in regard to melt pool lengths when using 10 and 20 samples, respectively. It is noted that the training outputs generally ``envelop'' the measurement data, with a few exceptions. This is a sign that the prior space is large enough to cover the true values of the calibration parameters $\bm{\theta}^{*}$. Increasing the bounds of the prior distributions may produce training outputs that fully envelop the experimental data, but can also make some uncertain inputs less physical, according to the model developers.

We used leave-one-out cross-validation (LOOCV) to test the accuracy of the GP surrogate models built with 5, 10, 15, 20, and 25 samples for each experimental condition. Table \ref{table:table5-Q2} shows the predictivity coefficient $Q_2$ for each condition. This coefficient $Q_2$ gives the percentage of the output variance explained by the GP surrogate model:
\begin{equation*}
	Q_2  =  1 - \frac{ \sum_{i=1}^{N_{\text{val}}} \left[ y^{\text{M}} ( \mathbf{x}^{(i)} ) - \mu_{\hat{y}} ( \mathbf{x}^{(i)} ) \right]^2 }{ \sum_{i=1}^{N_{\text{val}}} \left[ y^{\text{M}} \left( \mathbf{x}^{(i)} \right) - \overline{y^{\text{M}}} \right]^2 }
\end{equation*}
where $N_{\text{val}}$ is the size of the validation sample set, $y^{\text{M}} ( \mathbf{x}^{(i)} )$ is the output from the original model simulation, $\overline{y^{\text{M}}}$ is their empirical mean value, and $\mu_{\hat{y}} ( \mathbf{x}^{(i)} )$ is the predicted value using the GP metamodel. A $Q_2$ value close to 1.0 means that the model is accurate. In practical situations, a metamodel with a $Q_2$ value above 0.7 is often considered a satisfactory approximation of the original model \cite{wu2018inverse-part1}. It can be seen in Table \ref{table:table5-Q2} that, with more than 10 samples for each experiment test, the $Q_2$ values are above 0.95 for both responses.

\begin{table}[htbp]
	\centering
	\footnotesize
	\caption{Predictivity coefficient $Q_2$ for LOOCV of the GP model with different numbers of training samples.}
	\label{table:table5-Q2} 
	\begin{tabular}{ccc}
		\toprule 
		Number of samples  & Melt pool length  &  Melt pool depth    \\
		\midrule
		5   &  0.947  & 0.974  \\ 
		10  &  0.975  & 0.991  \\  
		15  &  0.972  & 0.991  \\ 
		20  &  0.976  & 0.984  \\  
		25  &  0.977  & 0.992  \\ 
		\bottomrule
	\end{tabular}
\end{table}

\begin{figure}[htb]
	\centering
	\includegraphics[width=0.85\textwidth]{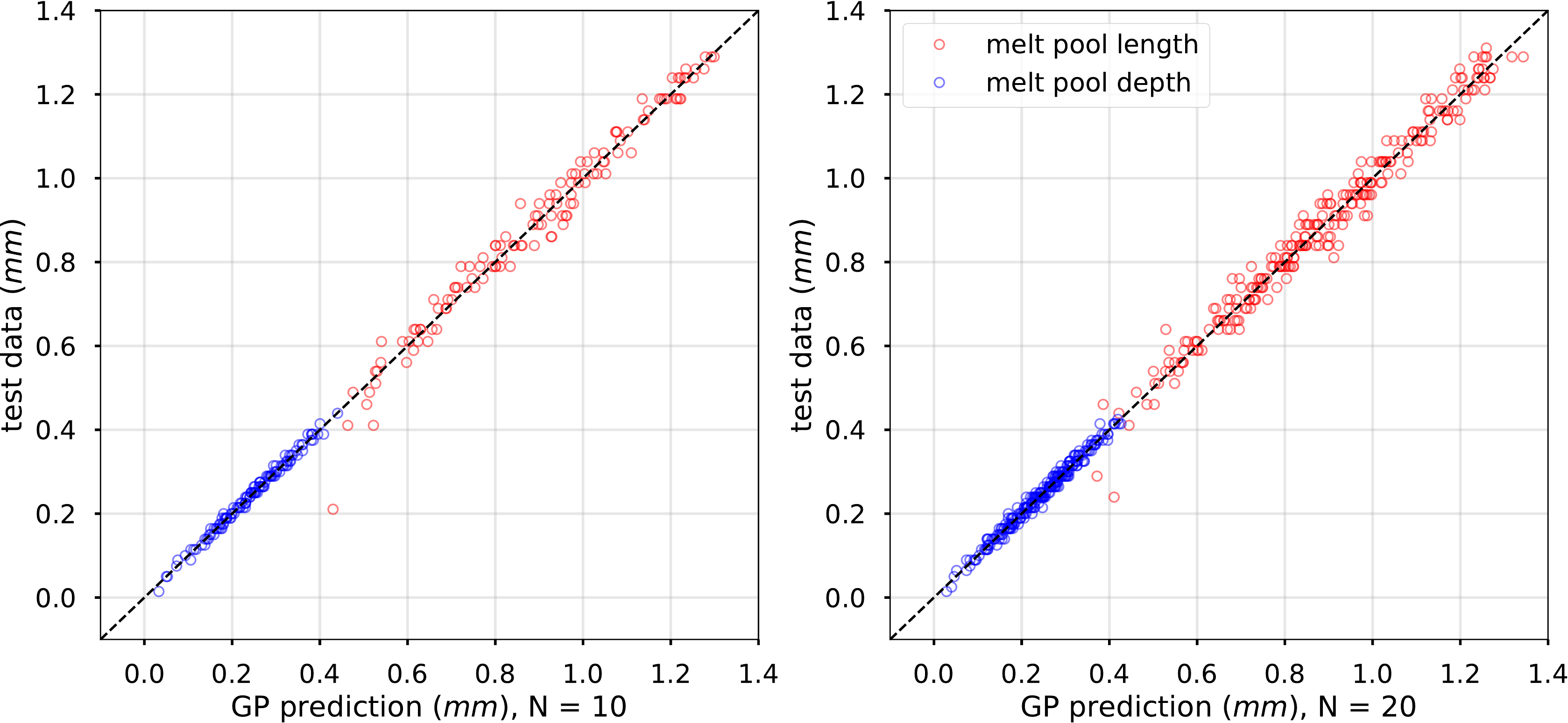}
	\caption[]{LOOCV of the GP surrogate models with 10 and 20 samples.}
	\label{figure:fig5-GP-Validation-LOOCV}
\end{figure}

Figures \ref{figure:fig5-GP-Validation-LOOCV} shows the LOOCV results (responses from the original model vs. responses from the GP surrogate model) for 10 and 20 samples, respectively. It can be seen that, with 10 samples for each experimental condition, the GP surrogate model accuracy is adequate, and no obvious improvement can be obtained by using more samples. Therefore, in the following sections, we chose the GP model trained by 10 samples for each measurement condition, with $13 \times 10 = 130$ training samples in total.

\subsection{Global sensitivity analysis}

One of the most concerning unresolved problems for inverse UQ is the ``lack of identifiability'' issue. Identifiability refers to the ability of inverse UQ to precisely distinguish among parameter uncertainty, model uncertainty, and data uncertainty when estimating the calibration parameters, since the discrepancy between the model and the experimental data is caused by all these uncertainties collectively. Even though model uncertainty is not considered in this work, inverse UQ may still be a slightly ill-posed problem. Previous work \cite{wu2019demonstration} has shown that, for a given calibration parameter to be statistically identifiable, it should be significant to at least one of the responses whose data is used for inverse UQ. Good identifiability cannot be achieved for a certain calibration parameter if it is not significant to any of the responses. Therefore, prior to inverse UQ, we will perform a sensitivity analysis (SA) using the validated GP surrogate model.

SA is a process of studying how the uncertainties in the QoIs can be apportioned to various random input parameters. SA provides a ranking of the input parameters according to their significance to QoIs. Global SA deals with the output variation caused by uncertain inputs over the whole domain globally. Popular global SA methods include regression-based, screening, and variance-based methods. In this work, we calculate the Pearson correlation coefficient (PCC), Spearman rank correlation coefficient (SRCC), and Sobol' indices as SA measures. Regression-based SA uses correlation coefficients, which are a numerical measure of some type of correlation that represents a statistical relationship between two variables. PCC measures the linear correlation between variables $X$ and $Y$:
\begin{equation}	 \label{equation:SA1-PCC}
	\rho_{X, Y}  =  \frac{\text{cov} \left( X, Y\right)}{\sigma_{X} \cdot \sigma_{Y}}  =  \frac{ \mathbb{E} \left[ ( X - \mu_{X} ) ( Y - \mu_{Y} )  \right] }{\sigma_{X} \cdot \sigma_{Y}}
\end{equation}

A rank correlation coefficient is a statistic that represents an ordinal association, the relationship between rankings of different ordinal variables or different rankings of the same variable. SRCC is a measure of how well the relationship between two variables can be described by a monotonic function. It is equal to the PCC for the rank values of those two variables. While PCC assesses linear relationships, SRCC assesses monotonic ones.

Variance-based SA methods mainly use the analysis of variance decomposition, which represents the variance of the output as a sum of the contributions of each input variable or combination thereof. Sobol' indices \cite{sobol2001global} \cite{saltelli2010variance} is the most popular method for variance-based SA. The first-order sensitivity coefficient, or first-order Sobol' indices, is defined as:
\begin{equation*}
	S_i = \frac{\text{Var}_{(X_i)} \left[ \mathbb{E}_{(\mathbf{X}_{\sim i})} \left( Y|X_i \right)  \right]}{\text{Var}(Y)}
\end{equation*}
where $Y$ and $\mathbf{X}$ are the model output and input, respectively. $\mathbf{X}_{\sim i}$ represents a list of all input factors except $X_i$. The inner expectation operator indicates that the mean of $Y$ is taken over all possible values of $\mathbf{X}_{\sim i}$ (given a fixed $X_i$), while the outer variance is taken over all possible values of $X_i$. $S_i$, also called the ``main effect,'' quantifies the variability in $Y$ caused solely by the uncertainty in $X_i$. Another kind of sensitivity index is defined as:
\begin{equation*}
	T_i = \frac{\mathbb{E}_{(\mathbf{X}_{\sim i})} \left[ \text{Var}_{(X_i)} \left( Y|\mathbf{X}_{\sim i} \right)  \right]}{\text{Var}(Y)} = 1 - \frac{\text{Var}_{(\mathbf{X}_{\sim i})} \left[ \mathbb{E}_{(X_i)} \left( Y|\mathbf{X}_{\sim i} \right)  \right]}{\text{Var}(Y)}
\end{equation*}
where $T_i$ measures the \textit{total effect} and is called the ``total sensitivity indices.'' Since $S_{\sim i} = {\text{Var}_{(\mathbf{X}_{\sim i})} \left[ \mathbb{E}_{(X_i)} \left( Y|\mathbf{X}_{\sim i} \right)  \right]} / {\text{Var}(Y)}$ can be understood as the first-order sensitivity indices, or main effect, of $\mathbf{X}_{\sim i}$, the expression $(T_i =  1 - S_{\sim i} )$ represents the portion of total variance $\text{Var}(Y)$ contributed by all input combinations that include $X_i$. Since $T_i$ includes the first-order effects of $X_i$ and higher-order effects of $X_i$ by interacting with other input factors, it is always larger than the main effect $S_i$.

\begin{figure}[ht]
	\centering
	\includegraphics[width=0.85\textwidth]{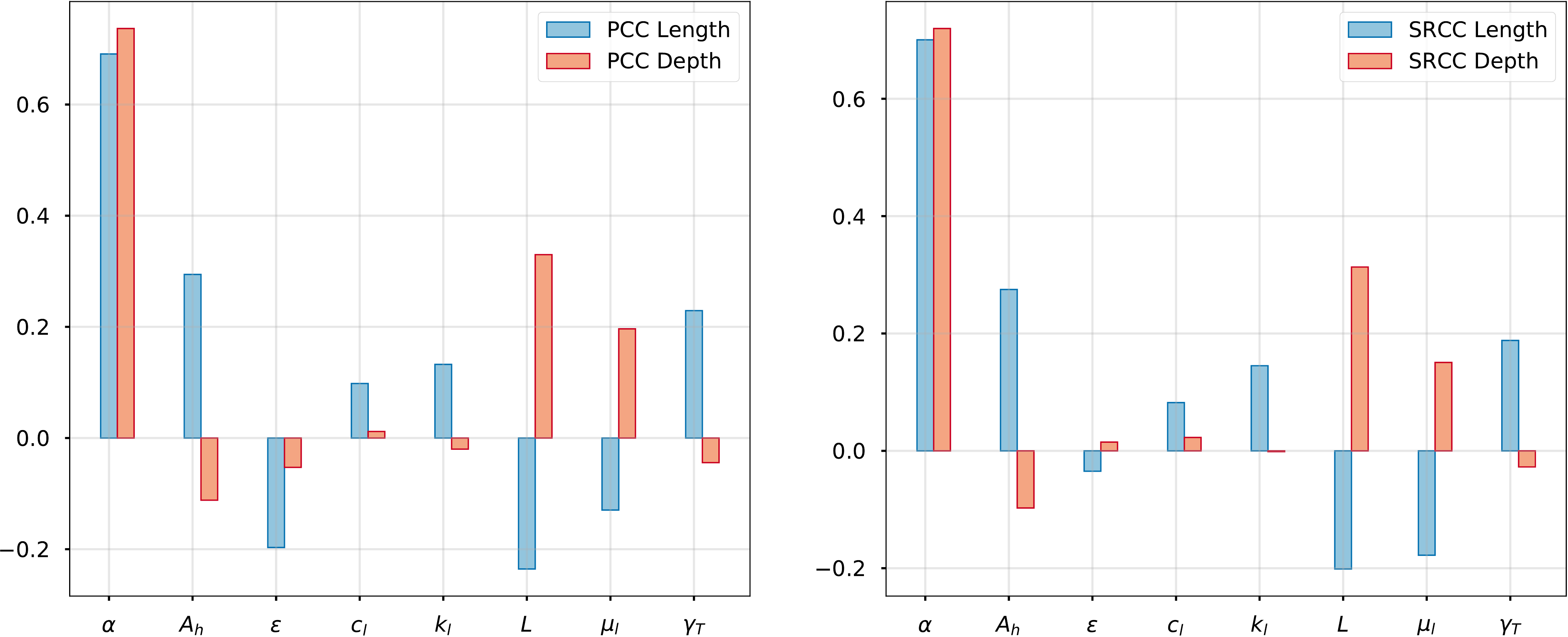}
	\caption[]{Comparison of PCC (left) and SRCC (right) results.}
	\label{figure:fig3-SA-CC}
\end{figure}

\begin{figure}[ht]
	\centering
	\includegraphics[width=0.85\textwidth]{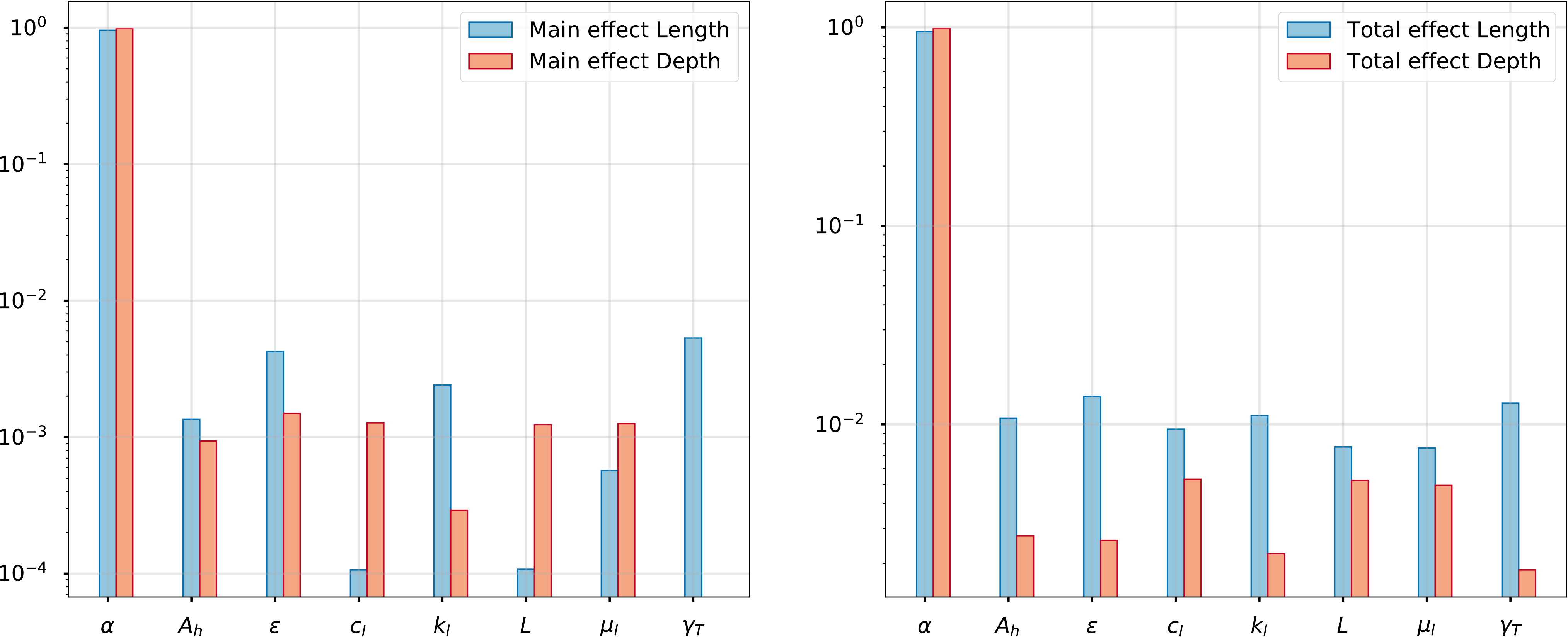}
	\caption[]{Comparison of main effect (left) and total effect (right) Sobol' indices.}
	\label{figure:fig3-SA-Sobol}
\end{figure}

The calculated values for PCC and SRCC are shown in Table \ref{table:table4-SA-CC}, while the results for the Sobol' indices are presented in Table \ref{table:table4-SA-Sobol}. Figures \ref{figure:fig3-SA-CC} and \ref{figure:fig3-SA-Sobol} provide a visualization of the SA results. It can be seen that PCC and SRCC produce very similar results. Both PCC/SRCC and the Sobol' indices show that the first parameter $\alpha$, which is the laser energy absorption coefficient, has dominant significance for both melt pool length and depth. However, Sobol' indices provides a more quantitative indicator, showing that $\alpha$ accounts for almost $100\%$ of the total variance in the responses. All the other calibration parameters are relatively trivial. Based on the findings in \cite{wu2019demonstration}, there is a risk that the uncertainties for all the remaining seven parameters cannot be quantified properly.

\begin{table}[ht]
	\centering
	\footnotesize
	\caption{Global SA results with PCC and SRCC.}
	\label{table:table4-SA-CC} 
	\begin{tabular}{crrrr}
		\toprule
		Parameters  &  \multicolumn{2}{c}{Melt Pool Length}  & \multicolumn{2}{c}{Melt Pool Depth} \\ 
		            &  PC  & SRCC  &  PCC  & SRCC  \\
		\midrule
		$\alpha$          & 0.644  & 0.654    & 0.757  &  0.738\\
		$A_h$             & 0.220  & 0.208    & -0.153 &  -0.164\\
		$\epsilon$        & -0.322 & -0.173   & -0.048 &  -0.006\\
		$c_l$             & 0.097  & 0.076    & 0.087  & 0.101\\
		$k_l$             & 0.070  & 0.090    & -0.037 &  -0.012\\
		$L$               & -0.206 & -0.180   & 0.302  & 0.271\\
		$\mu_l$           & -0.072 & -0.148   & 0.203  & 0.122\\
		$\gamma_T$        & 0.272  & 0.245    & -0.066 & -0.047  \\
		\bottomrule
	\end{tabular}
\end{table}

\begin{table}[ht]
	\centering
	\footnotesize
	\caption{Global SA results with Sobol' indices.}
	\label{table:table4-SA-Sobol} 
	\begin{tabular}{ccccc}
		\toprule
		Parameters  &  \multicolumn{2}{c}{Melt Pool Length}  & \multicolumn{2}{c}{Melt Pool Depth} \\ 
		            &  $S_i$  & $S_T$  &  $S_i$  &  $S_T$  \\
		\midrule
		$\alpha$          & 0.9571  & 0.9521    & 0.9857  & 0.9865 \\
		$A_h$             & 0.1351  & 0.0101    & 0.0009  & 0.0027 \\
		$\epsilon$        & 0.0042  & 0.0139    & 0.0015  & 0.0026 \\
		$c_l$             & 0.0001  & 0.0095    & 0.0013  & 0.0053 \\
		$k_l$             & 0.0024  & 0.0111    & 0.0003  & 0.0022 \\
		$L$               & 0.0001  & 0.0077    & 0.0012  & 0.0052 \\
		$\mu_l$           & 0.0006  & 0.0076    & 0.0013  & 0.0049 \\
		$\gamma_T$        & 0.0053  & 0.0128    & -0.0002 & 0.0019 \\
		\bottomrule
	\end{tabular}
\end{table}

\subsection{Inverse UQ results}

\begin{figure}[htbp]
	\centering
	\includegraphics[width=0.85\textwidth]{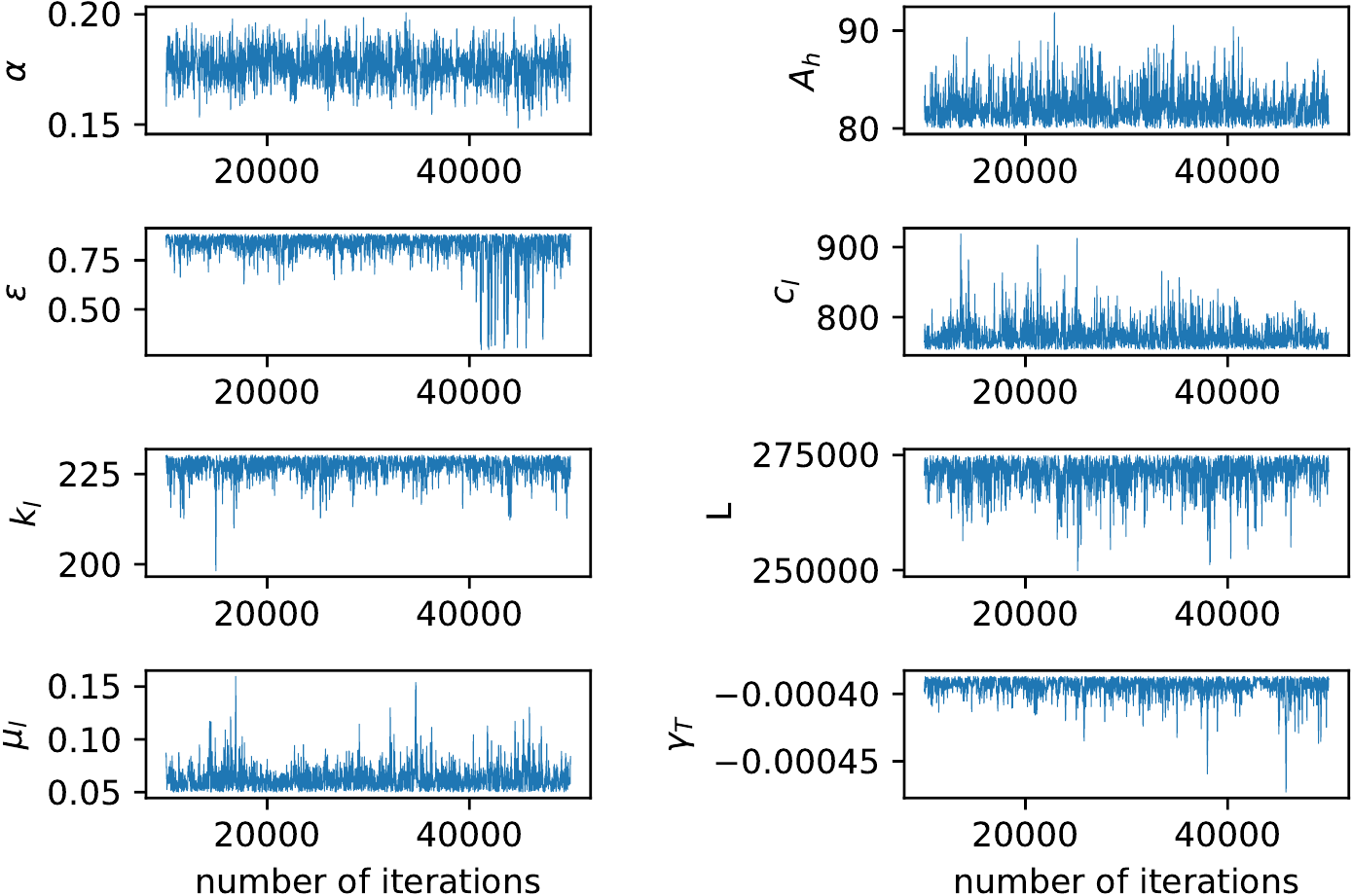}
	\caption[]{Mixing of posterior samples from MCMC sampling.}
	\label{figure:fig7-MCMC-chains}
\end{figure}

\begin{figure}[htbp]
	\centering
	\includegraphics[width=0.85\textwidth]{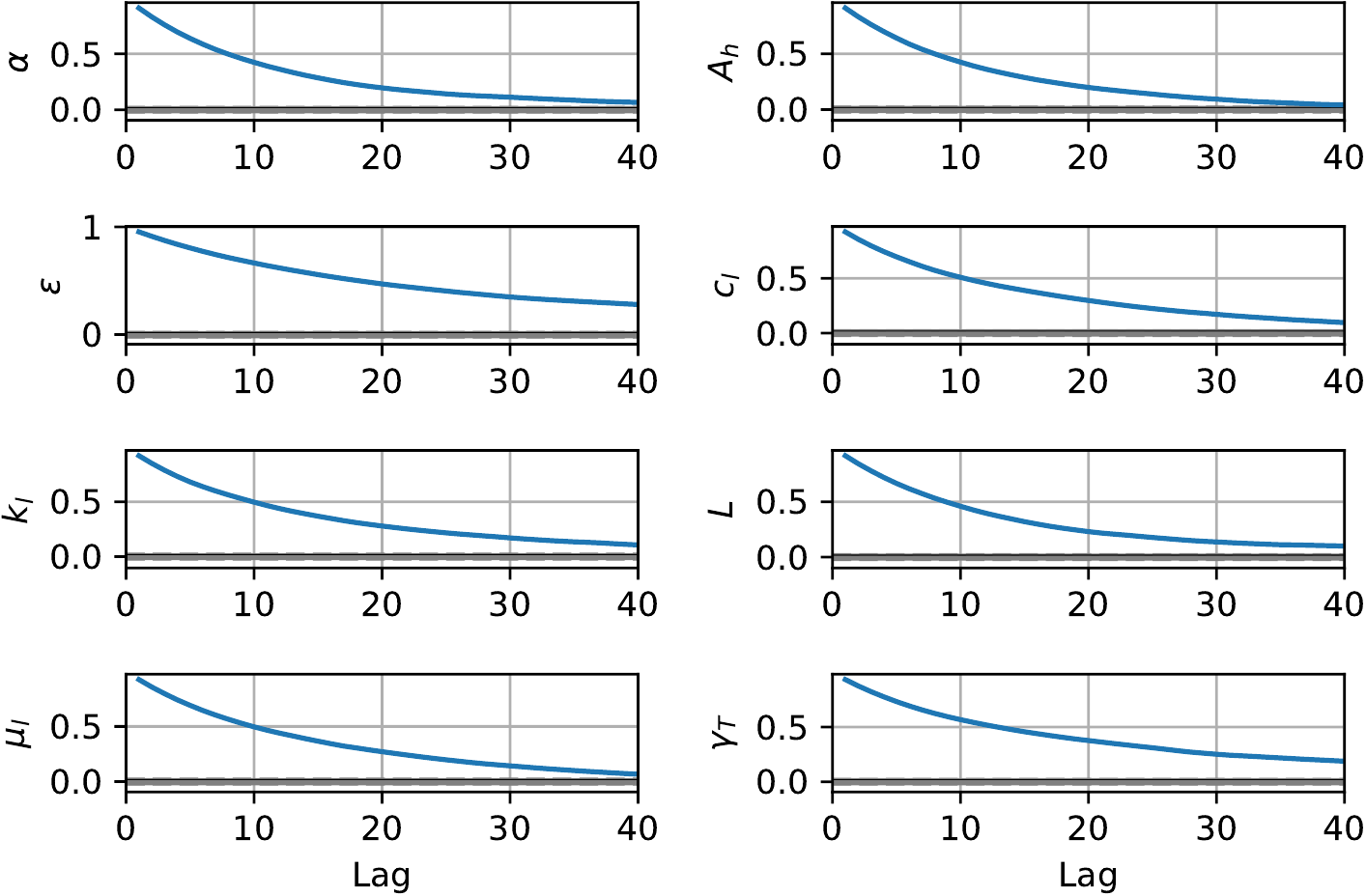}
	\caption[]{Auto-correlation plot for MCMC samples.}
	\label{figure:fig7-Autocorrelation}
\end{figure}

In the inverse UQ process, we used adaptive MCMC sampling \cite{andrieu2008tutorial} to generate posterior samples. A total of 50,000 MCMC samples were generated. The first 10,000 were abandoned as burn-in, since the MCMC process requires a certain number of steps to ``forget'' its initial location. Figure \ref{figure:fig7-MCMC-chains} shows the mixing of the remaining 40,000 MCMC samples. As expected, the mixing of the first parameter is very good, while that of the remaining seven parameters is not. Figure \ref{figure:fig7-Autocorrelation} shows the auto-correlation plot of the MCMC samples. We retained every 20$^{\text{th}}$ sample of the remaining 40,000 samples for the purpose of ``thinning,'' thus reducing the auto-correlations of the MCMC samples to a low level.

\begin{figure}[htbp]
	\centering
	\includegraphics[width=0.85\textwidth]{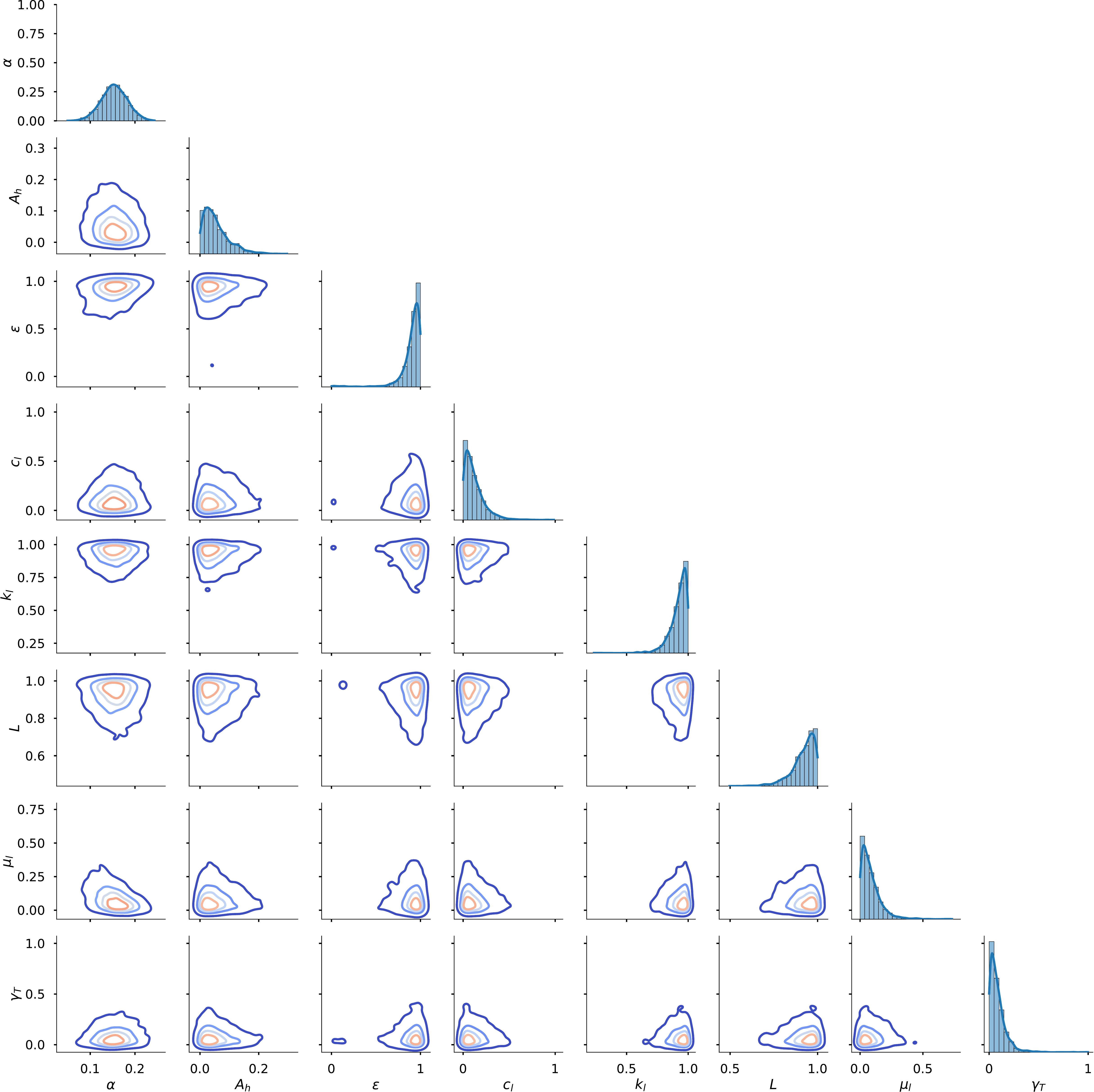}
	\caption[]{Posterior distributions based on MCMC samples, pair-wise joint densities (off-diagonal), and marginal densities (diagonal).}
	\label{figure:fig6-MCMC-samples}
\end{figure}

Figure \ref{figure:fig6-MCMC-samples} shows the posterior pair-wise joint densities (off-diagonal sub-figures) and marginal densities (diagonal sub-figures) for the eight calibration parameters, calculated using the remaining 2,000 samples. Besides the first parameter $\alpha$, whose shape is similar to that of normal distributions, all the remaining seven parameters are more or less concentrated to the upper/lower bounds. This indicates that the prior parameter ranges are not wide enough. However, based on the SA results, this can be due to the non-identifiability of the inverse UQ process, since these seven parameters have low significance for the two selected QoIs. Table \ref{table:table6-Posterior-Moments} presents the mean values, standard deviations, and 95\% confidence intervals based on the MCMC samples.

\begin{table}[htbp]
	\centering
	\footnotesize
	\caption{Posterior mean values, standard deviations, and 95$\%$ confidence intervals.}
	\label{table:table6-Posterior-Moments} 
	\begin{tabular}{cccc}
		\toprule
		Parameters  &  Mean values  &  Standard deviations  &  95$\%$ confidence intervals \\ 
		\midrule
		$\alpha$          & 0.18                    & 0.0078                  &   [0.161,0.192]      \\
		$A_h$             & 82.20                   & 1.75                    &   [80.11,86.74]      \\
		$\epsilon$        & 0.82                    & 0.078                   &   [0.622,0.883]      \\
		$c_l$             & 774.09                  & 19.36                   &   [745.29,823.05]    \\
		$k_l$             & 226.96                  & 2.92                    &   [219.64,227.72]    \\
		$L$               & $2.71{\times}10^5$      & $3.49{\times}10^3$      &   [$2.618{\times}10^5$,$2.749{\times}10^5$]    \\
		$\mu_l$           & 0.063                   & 0.012                   &   [0.050,0.094]    \\
		$\gamma_T$        & $-3.94{\times}10^{-4}$  &  $7.54{\times}10^{-6}$  &   [$-4.124{\times}10^{-4}$,$-3.872{\times}10^{-4}$]\\
		\bottomrule
	\end{tabular}
\end{table}

\subsection{Validation with the inverse UQ results}

\begin{figure}[htbp]
	\centering
	\includegraphics[width=0.95\textwidth]{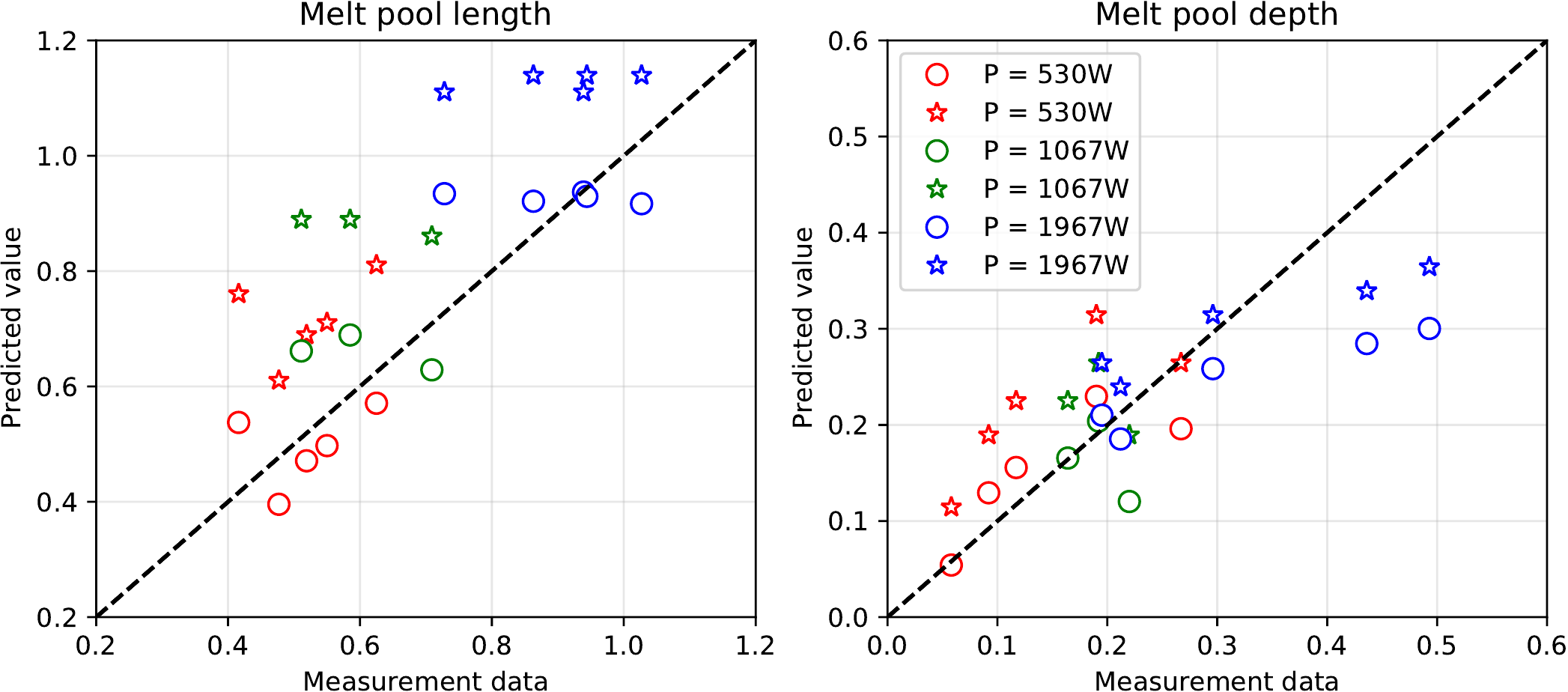}
	\caption[]{Comparison of simulation and experimental data for melt pool sizes. Circle: simulation using posterior mean values; Star: simulation using prior nominal values.}
	\label{figure:fig8-Simulation-at-Posterior-vs-Exp-Data}
\end{figure}

With the updated uncertainty information for the calibration parameters, we ran the original MOOSE-based melt pool model at the nominal values (i.e., mean values) of the posterior distributions. A comparison of simulation results based on prior nominal values, posterior mean values, and experimental data is shown in Figure \ref{figure:fig8-Simulation-at-Posterior-vs-Exp-Data}. It can be observed that, after inverse UQ, the agreement in melt pool length improved, while the agreement in melt pool depth remained good, except for two experiments in which the laser power was, 1967 W. However, it must be noted that this is not a legitimate ``validation,'' since we are comparing against the experimental data already used for inverse UQ. A more rigorous validation study should be performed with a new set of independent measurement data not used for inverse UQ.

\begin{table}[htbp]
	\centering
	\footnotesize
	\caption{Error in melt pool length and depth prediction (compared to experimental data) using prior nominal values and posterior mean values.}
	\label{table:table7-Validation-with-Posterior} 
	\begin{tabular}{ccccc}
		\toprule 
		Index  &  Prior       &  Prior         &  Posterior       &  Posterior  \\
		    &  length error (mm)  &  depth error (mm)  &  length error (mm)  &  depth error (mm)  \\
		\midrule
		1       &  0.186  & 0.124    & 0.054 & 0.039 \\ 
		2       &  0.344  & 0.002    & 0.121 & 0.071 \\  
		3       &  0.161  & 0.108    & 0.052 & 0.039 \\ 
		4       &  0.170  & 0.097    & 0.048 & 0.037 \\  
		5       &  0.134  & 0.056    & 0.081 & 0.004 \\ 
		6       &  0.304  & 0.072    & 0.104 & 0.012 \\ 
		7       &  0.378  & 0.061    & 0.150 & 0.002 \\ 
		8       &  0.152  & 0.031    & 0.080 & 0.099 \\ 
		9       &  0.172  & 0.129    & 0.002 & 0.193 \\ 
		10      &  0.382  & 0.097    & 0.206 & 0.151 \\ 
		11      &  0.195  & 0.018    & 0.014 & 0.037 \\ 
		12      &  0.276  & 0.069    & 0.058 & 0.015 \\ 
		13      &  0.112  & 0.027    & 0.011 & 0.027 \\
		Average &  0.228  & 0.068    & 0.083 & 0.056 \\
		\bottomrule
	\end{tabular}
\end{table}

Table \ref{table:table7-Validation-with-Posterior} shows the discrepancies between the experimental data for each experimental setting and the simulation results based on prior nominal values and posterior mean values. The index indicates different experimental settings, as listed in Table \ref{table:table3-Design-Variables-and-Exp-Data}. After Bayesian inverse UQ, the average error in melt pool length was reduced from 0.228 to 0.083 mm, while the average error in melt pool depth was reduced from 0.068 to 0.056 mm.

\section{CONCLUSIONS}

In this work, we quantified the uncertainties in the calibration parameters of a MOOSE-based melt pool model for AM simulation using experimental data. Inverse UQ aims to quantify uncertainties in calibration parameters while achieving consistency between code simulations and physical observations. Inverse UQ always captures the uncertainty in its estimates rather than merely determining point estimates of the best-fit input parameters. We employed a GP surrogate model to reduce the computational cost in MCMC sampling. Simulation results using the posterior uncertainties show much better agreement with experimental data, as compared to those using the prior nominal values. The resulting parameter uncertainties can be used to replace expert opinions in future uncertainty, sensitivity, and validation studies. Unfortunately, the inverse UQ process suffers from the non-identifiability issue, due to the fact that seven of the eight calibration parameters are not significant to the QoIs (i.e., melt pool length and depth). In the future, we will consider other types of melt pool model QoIs to properly quantify the uncertainties in these parameters.

Although AM shows considerable promise for nuclear fuel fabrication, the relation between the AM process parameters and the properties and performance of the AM-processed nuclear fuels has not yet been elucidated. In many cases, AM-processed products often exhibit lower performance than those produced via conventional methods \cite{cherry2015investigation} \cite{megahed2016metal} \cite{khairallah2016laser}. Sophisticated design and control of the AM process is critically challenging, especially in the selection of processing parameters (e.g., applied laser power, traveling speed, and scan style), an activity that determines the multiphase and multigrain mesoscale microstructures (e.g., crystallographic defects and textures)--in turn, greatly influencing the macroscale thermal and mechanical properties of fuel materials. Advanced computational M\&S techniques can provide insight into a rational design by addressing the basic science and engineering needs for the AM process, leading to a mechanistic understanding of processing-structure-property relationships, which are fundamental to the design and development of innovative fabrication techniques for nuclear fuels. Therefore, future research is needed into AM process optimization in order to determine an optimal set of AM process parameters.

\section*{ACKNOWLEDGEMENT}

This work was supported through INL's Laboratory Directed Research \& Development (LDRD) Program under DOE Idaho Operations Office Contract DE-AC07-05ID14517. This research made use of INL's high-performance computing resources, which are supported by the Office of Nuclear Energy of the U.S. Department of Energy and the Nuclear Science User Facilities under Contract No. DE-AC07-05ID14517.

\bibliography{./Journal_IUQ_Melt_Pool.bib}

\end{document}